\documentclass[conference]{IEEEtran}

\usepackage{mathrsfs}
\usepackage{hologo}

\ifCLASSINFOpdf
 
\else
  
\fi

\usepackage{amsthm}
\usepackage{makeidx}
\usepackage{examples}
\usepackage[pdftex]{graphicx}  
\usepackage{amssymb}
\usepackage{graphicx}
\usepackage{algorithm}
\usepackage{epstopdf}
\usepackage{color}
\usepackage{url}
\graphicspath{{../eps/}{../ps/}}
\usepackage{psfrag}    
\usepackage{algorithm}
\usepackage{algorithmic}
\usepackage{amsmath}

\IEEEoverridecommandlockouts
\begin{document}

\title{3DNA$:$ A Tool for DNA Sculpting}

\author{
\IEEEauthorblockN{Shikhar Kumar Gupta, Foram Joshi, Dixita Limbachiya and Manish K Gupta\\
Laboratory of Natural Information Processing,\\ 
Dhirubhai Ambani Institute of Information and Communication Technology\\
	Email$:$ shikhar\_gupta@daiict.ac.in, foram\_joshi@daiict.ac.in, dlimbachiya@acm.org, m.k.gupta@ieee.org\\
}}

\maketitle

\begin{abstract}
DNA self$-$assembly is a robust and programmable approach for building structures at nanoscale. Researchers around the world have proposed and implemented different techniques to build two dimensional and three dimensional nano structures. One such technique involves the implementation of DNA Bricks \cite{Ke30112012}, proposed by Ke et al., $2012$ to create complex three$-$dimensional $(3$D$)$ structures. Modeling these DNA nano structures can prove to be a cumbersome and tedious task. Exploiting the programmability of base$-$pairing to produce self$-$assembling custom shapes, we present a software suite $3$DNA, which can be used for modeling, editing and visualizing such complex structures. $3$DNA is an open source software which works on the simple and modular self assembly of DNA Bricks, offering a more intuitive  better approach for constructing $3$D shapes. Apart from modeling and envisaging shapes through a simple graphical user interface, $3$DNA also supports an integrated random sequence generator that generates DNA sequences corresponding to the designed model. The software is available at www.guptalab.org$/$$3$dna
\end{abstract}


\IEEEpeerreviewmaketitle
\section{Introduction}
Construction of nano devices and nano$-$structures using the approach on self assembly is one of the most engrossing and upcoming field of research in DNA nanotechnology \cite{seeman1983design}.  Along with static structures, various dynamic models like molecular switches, DNA walkers, DNA robots, molecular circuits \cite{omabegho2009bipedal}, \cite{seelig2006enzyme}, \cite{zhang2011dna} are being developed. These structures are built by designing the DNA sequences in a specific manner that enforces the DNA to bind with its complementary base pairs. In the forefront, researchers have made attempts to built structures arbitrarily, but the field received a boost with the introducion of the idea of DNA origami by Paul Rothemund, in which a scaffold DNA sequence $($which is often a viral genomic DNA$)$, can be folded into a desired fashion by using synthetic staples strands \cite{rothemund2005design}, \cite{doi:10.1021/nn800215j}, \cite{Han15042011}, \cite{Castro_Kilchherr_Kim_Shiao_Wauer_Wortmann_Bathe_Dietz_2011}, \cite{Rothemund_2006}, \cite{WIN1}, \cite{winfree1998design}. There are list of different $2$D and $3$D nano structures \cite{park2006finite}, \cite{Dietz_Douglas_Shih_2009}, \cite{Han15042011}, \cite{Nangreave2010608}, \cite{kirigami10} built by using various DNA self assembly approaches. In subsequent years, Peng Yin et al., gave rise to an approach of modular self$-$assembly which employs DNA tiles. These tiles are single stranded tiles$($SSTs$)$ and they ultimately assemble into finite $2$D shapes \cite{wei2012complex}. This technique has paved the way for an efficient, simple and systematic approach to self assembly. Following the method of DNA tiles, Ke et al. extended the idea to DNA bricks \cite{Ke30112012}, which allows the construction of $3$D shapes. 

To aid the various applications of building these $3$D DNA structures, $3$DNA implements the concept of modular assembly of DNA bricks to construct $3$D shapes. $3$DNA can be employed to minimize the time$-$consuming and error$-$prone task of designing DNA sequences to model these formations. The software provides a $3$D molecular canvas interface where the user can model/design complex DNA structures. It also includes a sequence generator which computes the DNA sequences corresponding to the structure.

The sequences generated by the software self$-$assemble in one step annealing reactions into prescribed $3$D shapes. The software interface includes a $3$D molecular canvas of varying dimensions composed of several molecular pixels, which ultimately represent DNA bricks. By deselecting pixels from the molecular canvas we have been able to create different shapes of varying dimensions. Using $3$DNA we have designed complex shapes with intricate interior cavities. The use of $3$DNA significantly reduces the effort required to design $3$D DNA structures.

This paper is organized as follows. Section $2$ describes an outline of GUI and section $3$ provide a detailed description of the software functionality. Algorithms for strand construction and workflow are defined in section $4$ followed by detailed analysis of a sample case study in section $5$  and finally conclusion in section $6$.  Section $7$ provides a link for downloading the software and related materials.

\section{GRAPHICAL USER INTERFACE}
\begin{figure}
\includegraphics[scale=.40]{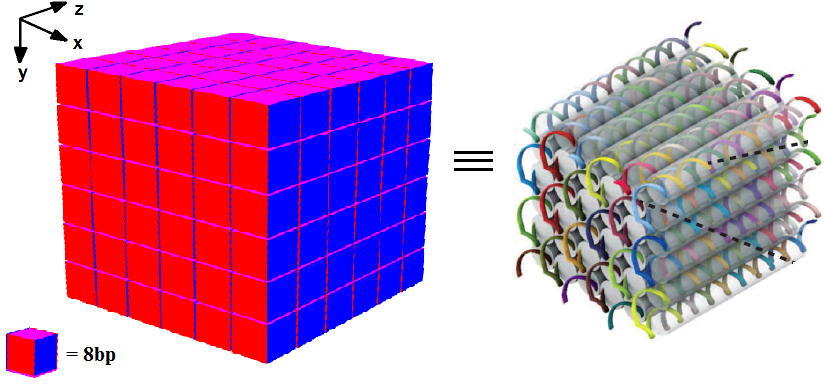}
\caption{A model that depicts a $6$H $\times$ $6$H $\times$ $48$B $3$D cuboid molecular canvas. Image Credit: \cite{Ke30112012}}
\label{canvas}
\end{figure}

The graphical user interface $($GUI$)$ for $3$DNA has been developed to enable the user to edit/model and visualize complex $3$D shapes on the molecular canvas $($as shown in Fig. \ref{canvas}$)$. By rendering the $3$D canvas, the user can edit molecular pixels and envisage the shape drawn at nano scale. 

\subsection{Molecular Canvas}

The customizable Java$3$D based molecular canvas can be viewed as a block$($cube$)$ of molecular pixels, each representing an 8$-$nt duplex and has a dimension of approximately $2$.$5$ by $2$.$5$ by $2$.$7$nm. The convention for setting the dimensions for the 3D molecular canvas are$:$ height and width in terms of DNA helices along the x \& y axis respectively, depth in terms of DNA base$-$pairs $($multiples of $8$$)$ along the z$-$axis.
This $3$D model contains the positional information of each 8$-$bp duplex in every pixel, which can be removed independently.  

\subsection{Canvas Control Panel}
The user interface provides the ability to control and fiddle  with the $3$D environment by changing the camera view using the Canvas Control Panel. The panel is equipped with zoom in, zoom out and other directional buttons for viewing the canvas from different angles and a better visualization of the structure.

\section{FUNCTIONALITY}
\begin{figure}
\includegraphics[scale=.40]{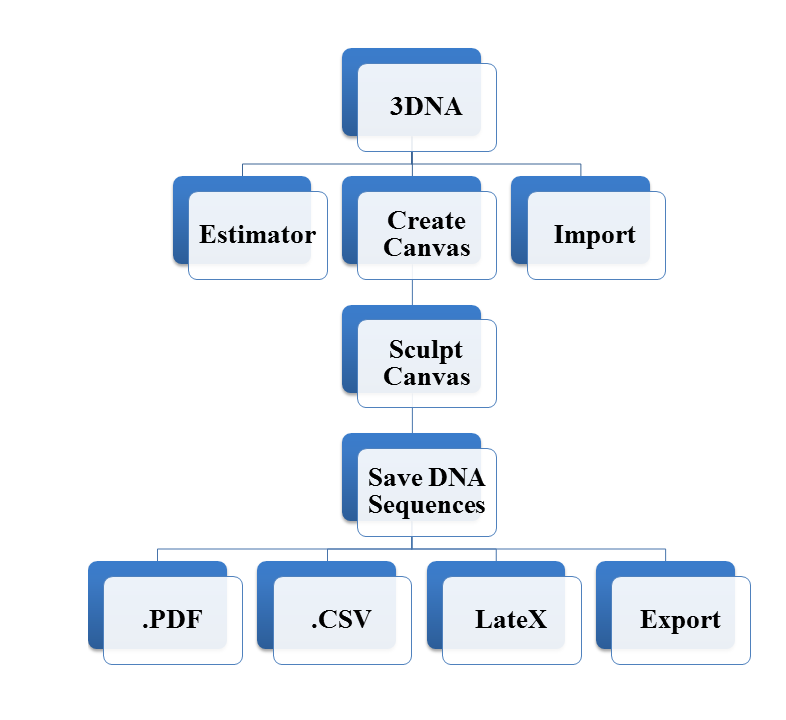}
\caption{A Flowchart depicting the main functionality and modules of $3$DNA}
\label{functionality}
\end{figure}

The following subsections are aimed to give a brief idea about the functionality $($Fig. \ref{functionality}$)$ of the software as a whole. The main functional features of the software are its input modules$($the \textbf{$3$D canvas}$)$ and \textbf{import}, its computational modules as well as its output modules which include \textbf{saving sequences} into files and \textbf{export}. 

\subsection{Creating a new canvas}
\begin{figure}
\includegraphics[scale=.60]{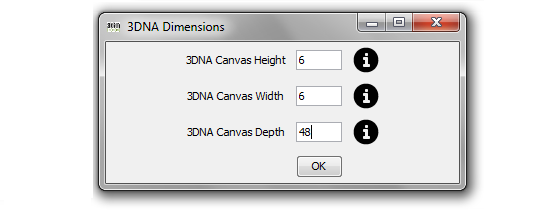}
\caption{User prompt to enter the dimensions of the canvas manually}
\label{prompt}
\end{figure}
The $3$D molecular canvas can be accessed by selecting the \textbf{New Canvas} button from the menu bar. On initiating the molecular canvas, the user will be prompted to enter the dimensions i.e. the height $($DNA helices$)$, width $($DNA helices$)$ and depth $($base pairs and a multiple of $8$$)$ of the $3$D canvas $($Fig. \ref{prompt}$)$. After entering the dimensions a custom Java$3$D based molecular canvas appears on the screen with the specified dimensions.

\subsection{Creating structures}

\begin{figure}
\includegraphics[scale=.30]{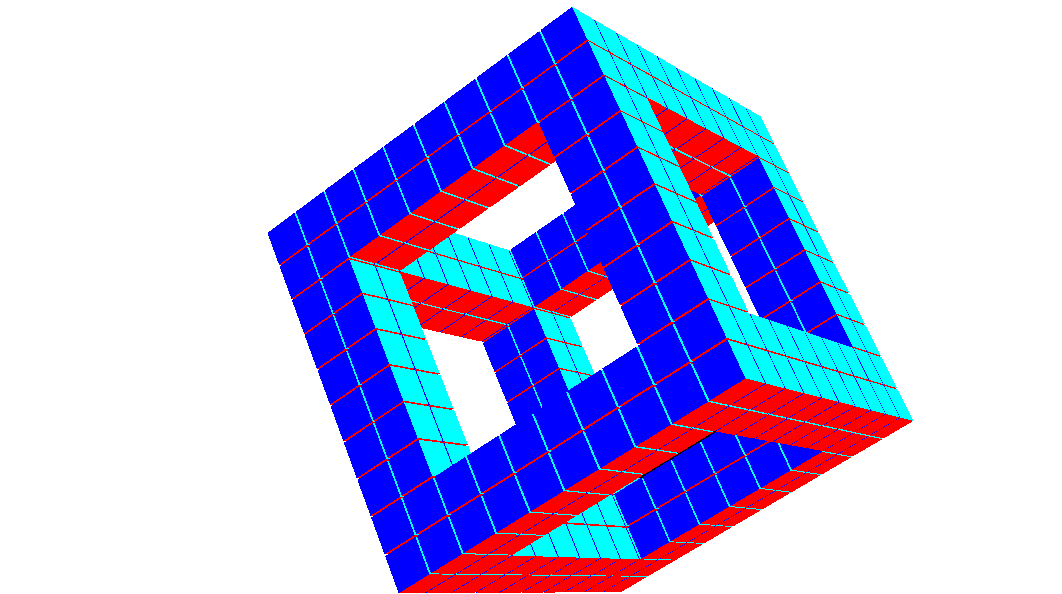}
\caption{Sample image of a $10$H $\times$ $10$H $\times$ $80$B $3$D hollow cube sculpted using $3$DNA molecular canvas}
\label{sample}
\end{figure}

The smallest unit of the $3$D molecular canvas is the molecular pixel, which represents $8$$-$bp. Each of these pixels combine according to the software prescribed algorithm into full$($$32$$-$nt$)$ or half$($$16$$-$nt$)$ bricks and can be removed independently. To create structures, the canvas allows the user to freely deselect a molecular pixel by simply clicking on it, thereby depicting sculpting of the DNA block. Fig. \ref{sample} displays a sample structure of a $3$D hollow cube sculpted by removing the unwanted pixels from an original cube of dimensions $10$H $\times$ $10$H $\times$ $80$B $($more details on the same can be found in section $5$$)$.  

\subsection{Visualization of the structure}
Visualization can be used to gain deeper insights on the structural bindings of the sculpture. It shows how the base pairs interact and associate with each other to envisage the shape at nanoscale.

\subsection{Analysis of sequences}
\begin{figure}
\includegraphics[scale=.40]{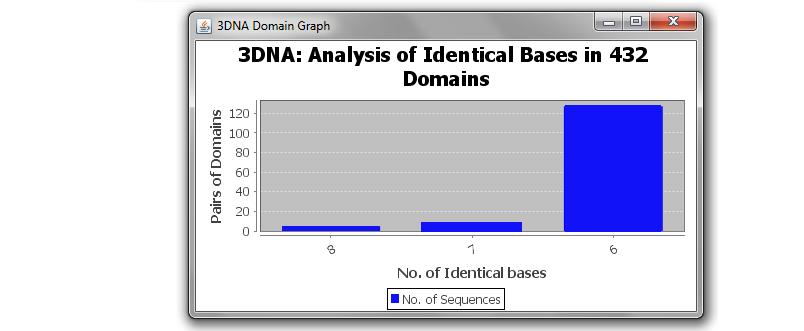}
\caption{A graphical analysis of a cube of 6H $\times$ $6$H $\times$ $48$B showing number of pairs of $8$$-$base domains that contain $8$, $7$, or $6$ identical bases among $432$ the domains in the structure }
\label{analysis}
\end{figure}

The final set of sequences generated by the software can be analyzed by the \textbf{Graphical Analysis} option on the menu bar. It generates a statistical analysis of the number of pairs of $8$$-$base domains that contain $8$, $7$, or $6$ identical bases among all the domains in the final sculpted structure. Fig. \ref{analysis} demonstrates this functionality for a $6$H $\times$ $6$H $\times$ $48$B cube. It indicates the number of $8$, $7$, or $6$ identical bases in the entire group of $432$ domains required make up the targeted structure. This analysis is useful to understand the nature of single stranded DNA sequences when conducting experiments. One can use this analysis to measure the stability and tolerance of identical bases in the domain sequences of the actual self$-$assembled structures.

\subsection{OUTPUT}
When a user sculpts the targeted $3$D structure on the canvas, the corresponding DNA sequences are generated. $3$DNA allows the user to save the DNA sequences in a variety of formats$:$

\begin{itemize}
\item Save as .pdf file$:$ DNA sequences designed by the software for creating the prescribed shape on the $3$D molecular canvas can be saved in a .pdf format by accessing the \textbf{Save as .pdf} button. The sequences pertaining to the full and half bricks are specified in a table along with the corresponding molecular pixels $($voxels$)$ to which they assigned. Each file saved in this format contains a unique bar code for identification. 

\item Save as .csv file$:$ The final DNA sequences along with the voxel coordinates generated by the software are saved into a .csv file by accessing the \textbf{Save as .csv} button.

\item Save as \hologo{LaTeX} file$:$ To save the final sequences into a \hologo{LaTeX} file, one can simply click on the  the \textbf{Save as \hologo{LaTeX} file} button.

\item Export and Import$:$ To provide flexibility and ability to open and edit an existing project, the user may export a current project. By availing the \textbf{Export} button, the current project is saved as a .$3$dna file. An existing project which has been exported previously can be opened and modified/viewed through the Import functionality by clicking on the \textbf{Import} button. Any existing project with .$3$dna extension can be imported.
\end{itemize}

\subsection{Cost Estimator}
It calculates the cost of experiment$($USD$)$ based on the total number of nucleotides. 

\section{ALGORITHMS AND WORKFLOW}
The following section expounds upon the methods and algorithms used during strand design and modelling of sequences. It also covers the workflow of the software starting from creating the basic canvas to obtaining the final sequences and the explanation of the intermediate steps in its process.

\subsubsection{Domain Sequence Generation}
Each of the four 8$-$nt domains of the canonical full brick and two 8$-$nt domains of half bricks are designed by imposing the following constraints on completely random assignment of base pairs$($A$-$T, G$-$C$)$. $($These constraints may or may not apply to the complimentary strands of domains$)$$:$
\begin{itemize}
\item No repetition of nucleotide beyond a length of four 
\item GC content ranging from $40$$-$$60$\%
\item Hamming distance of $6$ between all domains is maintained
\end{itemize}

\subsubsection{Molecular Pixel to Brick Sequence Generation}
In order to understand how each molecular pixel on the canvas is ultimately mapped to its corresponding brick sequence, it is important to understand the theory of the DNA single stranded bricks, their types, orientation and modeling$:$
\begin{itemize}
\item Full brick$:$ A full brick $($Fig. \ref{fullBrick}$)$ is 32$-$nt and is conceptualized as four consecutive 8$-$nt domains. Each DNA brick has a unique nucleotide sequence. An identical shape is assumed by all the DNA bricks upon the target formation, with the two 16$-$nt antiparallel helices joined by a single phosphate linkage in the center.

\begin{figure}
\includegraphics[scale=.30]{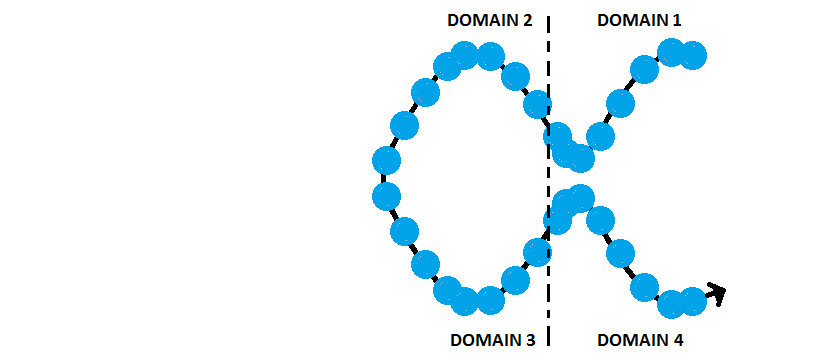}
\caption{A depiction of a full brick and its four domains through a helical single$-$stranded structure}
\label{fullBrick}
\end{figure}

\item Half brick$:$ The half brick$($Fig. \ref{halfBrick}$)$ is a bisection of the full brick representing a single helix with two domains.

\begin{figure}
\includegraphics[scale=.30]{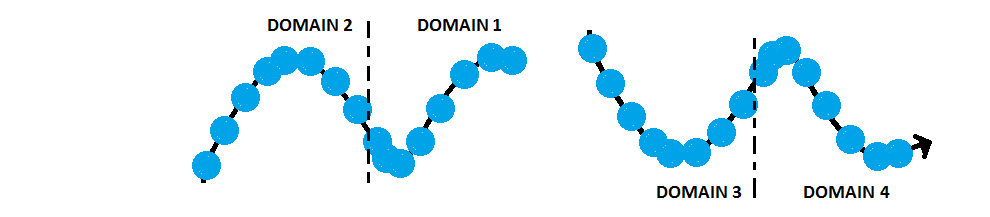}
\caption{A depiction of a half brick and its domains through a helical single$-$stranded structure}
\label{halfBrick}
\end{figure}

\item Orientation and modeling of bricks$:$ As introduced by Ke et al. the bricks adopt the LEGO modeling \cite{Ke30112012} of DNA. Each brick can be conceptualized as a LEGO cube$:$ the domains $1$ and $4$ form the protruding ends and domains $2$ and $3$ form the backbone of the LEGO cube. The bricks adopt one of four orientations$-$ north, west, south or east and therefore must be either horizontal or vertical. A full brick attaches to four immediate neighbors, which are complementary in sequence and perpendicular in orientation with it $($Fig. \ref{bricks}$)$.

\begin{figure}
\includegraphics[scale=.20]{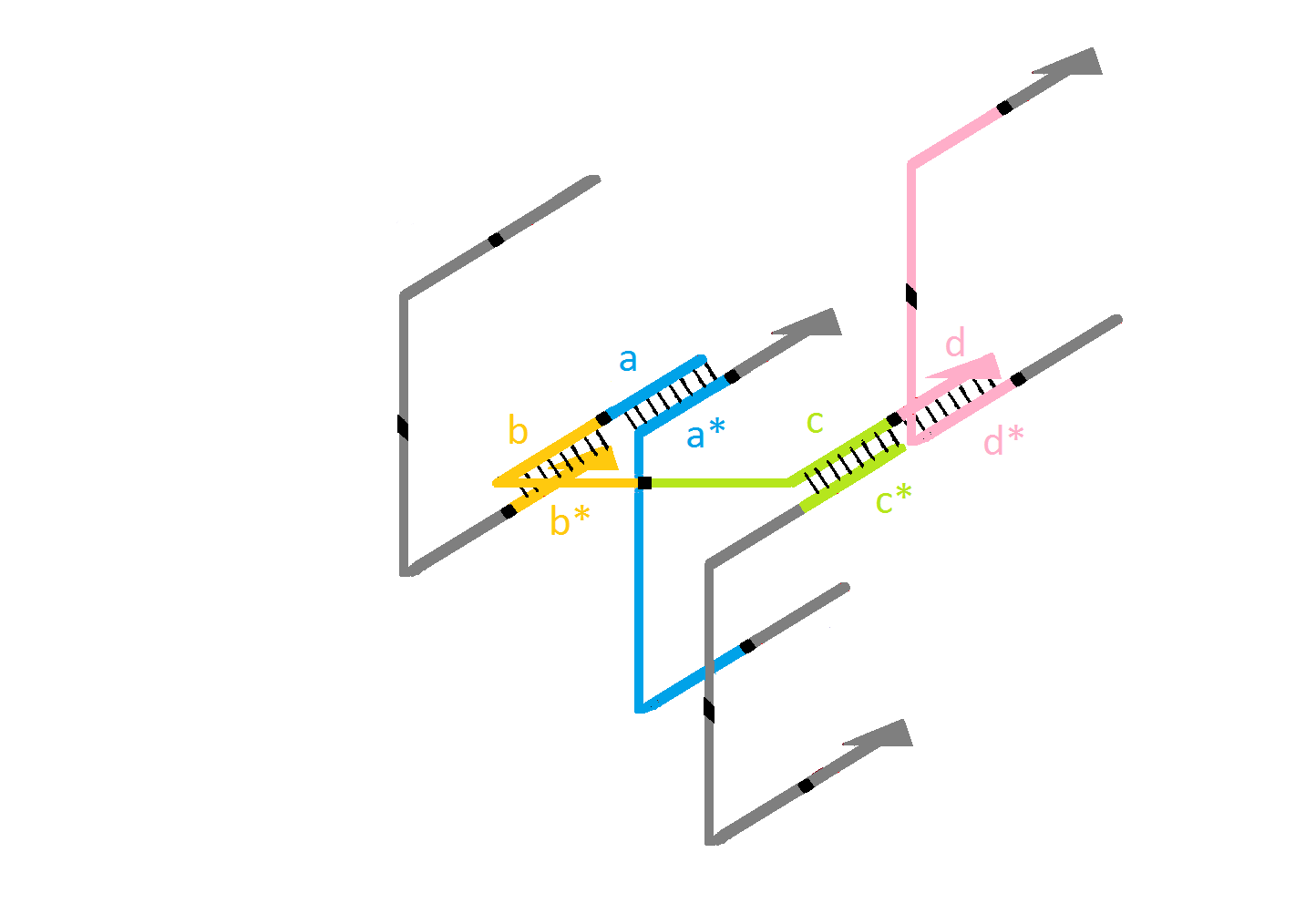}
\caption{Interaction of a full brick with its neighbors, showing the way complementary strands of the domains $($marked by the same color$)$ bind to each other}
\label{bricks}
\end{figure}
\end{itemize}

\subsection{WORKFLOW}
\begin{figure}
\includegraphics[scale=.50]{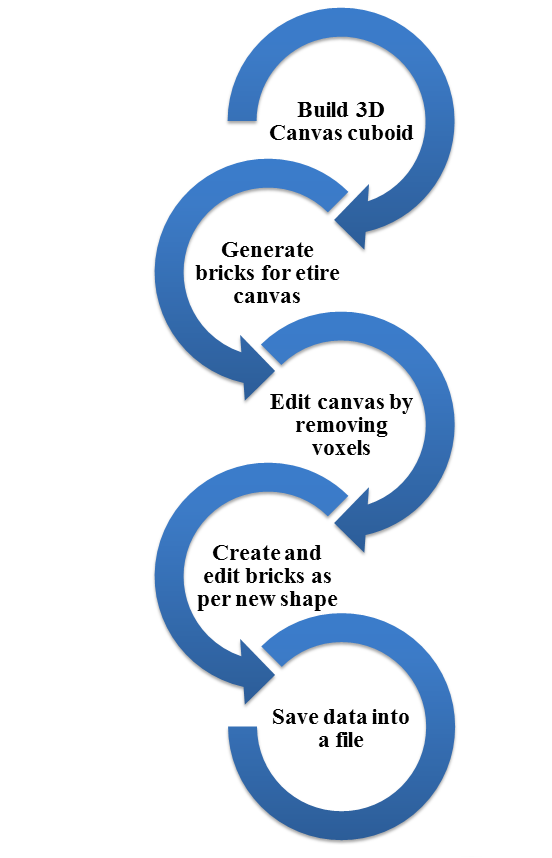}
\caption{A depiction of the workflow in $3$DNA}
\label{workflow}
\end{figure}

The complete workflow starting from creation of canvas to saving the subsequent sequences into files is illustrated in Fig. \ref{workflow}. $3$DNA is designed to obey the following steps $($workflow$)$ for every target structure$:$
\begin{enumerate}
\item Build $3$D canvas of preferred dimensions.
\item Design sequences for all the full and half bricks comprising the entire canvas.
\item Process the sculpted shape after user has removed the unwanted molecular pixels $($voxels$)$ and select the required subset of bricks needed for the structure. 
\item Modify the acquired subset of sequences and assign protector bricks $($unpaired strands composed of $8$ continuous thymidines to prevent unwanted interactions between exposed single$-$stranded domains$)$ and boundary bricks $($$48$$-$nt strands formed by merging a $32$-nt full brick and a $16$$-$nt half brick$)$. 
\item Save the final sequences pertaining to the targeted $3$D formation.
\end{enumerate}  

The final structures are self$-$assembled into their target shapes in one$-$step reactions. $($DNA brick self$-$assembly is the process by which DNA strands behave as LEGO bricks and adopt a defined arrangement without guidance or management from an external source.$)$

\section{Sample Structures}

Using $3$DNA, we have been able to sculpt various shapes of different sizes and ratios. For example, an initial cube of dimensions $8$H $\times$ $8$H $\times$ $64$B, which measures approximately $20$nm $\times$ $20$nm $\times$$21.6$nm was sculpted into shapes like a hollow cube $($Fig \ref{sample}$)$ and a gear $($Fig \ref{sample2}$)$. The canvas thus consists of $8$ $\times$ $8$ $\times$ $8$ molecular pixels, accounting for a total of $1024$ Domains, which constitute a sum of $288$ single strands of DNA. These $288$ strands and further classified into full bricks and half bricks which are $224$ and $64$ in number. The total number of nucleotides $($A,T,G,C$)$ required to build this cube is $8192$. Details of the sample files generated for the hollow cube and gear are mentioned below in sections A and B respectively.

\subsection{Hollow Cube}

The following inferences can be made about the hollow cube $($as shown in Fig \ref{sample}$)$$:$
\begin{itemize}
\item The hollow cube is obtained by deselecting $256$ of the $512$ molecular pixels from the canvas.
\item The sculpture has the same dimensions i.e., approximately $20$nm $\times$ $20$nm $\times$ $21.6$nm. This is due to the fact that the deselected pixels are all internal pixels.  
\item The hollow cube contains $512$ domains, which constitute a total of $168$ strands, out of which $80$ are half bricks and the remaining $88$ are full bricks. 
\item This structure requires a total of $4096$ nucleotides for its formation.
\item Assuming that the cost per base is U.S. Dollar $0$.$004$ , the total cost of the experiment would be $16$.$3$Dollars.
\end{itemize}

\subsection{$3$D Gear}
\begin{figure}
\includegraphics[scale=.40]{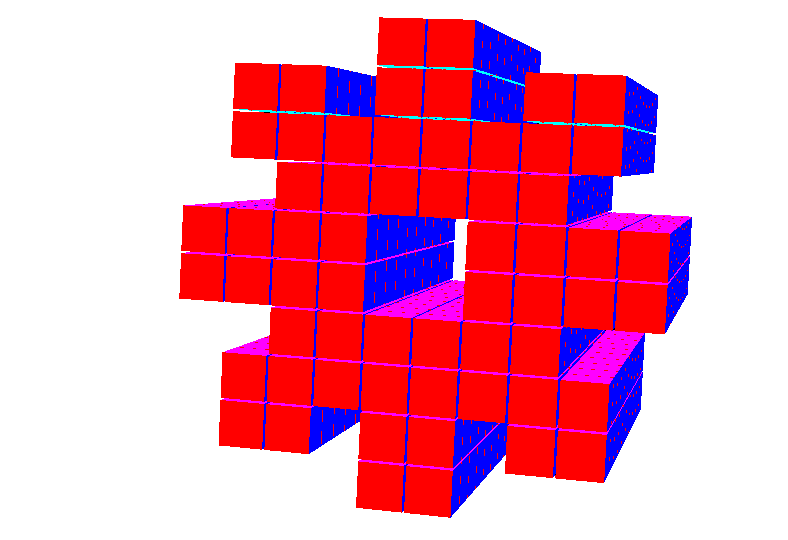}
\caption{Sample image of a $3$D gear sculpted from a $10$H $\times$ $10$H $\times$ $80$B molecular canvas}
\label{sample2}
\end{figure}

\begin{figure}
\includegraphics[scale=.40]{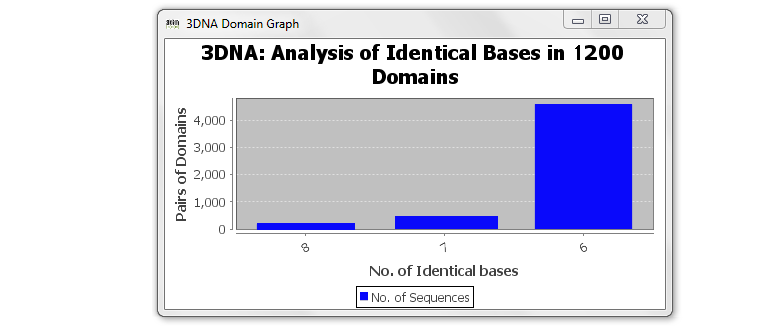}
\caption{A graphical analysis of a cube of $10$H $\times$ $10$H $\times$ $80$B showing number of pairs of $8$$-$base domains that contain $8$, $7$, or $6$ identical bases among $1200$ the domains in the gear }
\label{gridAnalysis}
\end{figure}

The following inferences can be made about the gear $($as shown in Fig \ref{sample2}$)$$:$
\begin{itemize}
\item The gear is sculpted form a $10$H $\times$ $10$H $\times$ $80$B  canvas. 
\item The gear is obtained by deselecting $440$ of the $1000$ molecular pixels from the canvas.
\item The sculpture has the same dimensions as the initial canvas i.e., approximately $25$nm $\times$ $25$nm $\times$ $27$nm.  
\item The gear contains $1200$ domains, which constitute a total of $380$ strands containing both full and half bricks. 
\item This structure requires a total of $9600$ nucleotides for its formation.
\item Fig \ref{gridAnalysis} shows a statistical estimation of the similarity of $8$, $7$ and $6$ bases among the domains. It can be used measure the stability and tolerance of identical bases in the domain sequences of the actual self$-$assembled structures. 
\item Assuming that the cost per base is U.S. Dollar $0$.$004$  , the total cost of the experiment would be Dollars $38$.$4$.
\end{itemize}

\section{Conclusion}
$3$DNA has been developed for the sole$-$purpose of generating a user$-$friendly, interactive environment for users to envisage their DNA structures, and get the actual DNA sequences required to make the physical formations. With the feature of edit dimensions, user can scale the shape in desire dimension and can view it with different orientations. Thus the output sequences can be experimentally used to make the nanoscale architectures with specified brick design. In the future, we expect to enhance the functionality of the software and enable the user to draw more complex structures.

\section{SOFTWARE AVAILABILITY}
The software source code, user manual, and supplementary materials can be downloaded from$:$\\ 
http$:$//www.guptalab.org/$3$dna.

\section{ACKNOWLEDGMENT}
The authors would like to thank Bruno Lowagie $($http$://$itexpdf.com$)$ whose libraries have been used in the project for generating bar codes and PDF and Nick Roach for providing us with a set of elegant icons thoroughly used in the software $($http$://$www.elegantthemes.com$)$.
\bibliography{3dnaref}
\bibliographystyle{IEEEtran} 

\end{document}